\documentclass[aps,prl,twocolumn,twoside,superscriptaddress,nofootinbib,preprintnumbers]{revtex4-2}

\usepackage[bookmarks=false]{hyperref}
\usepackage{graphicx}
\usepackage{amsmath}
\usepackage{amsfonts}
\usepackage{mathrsfs}
\usepackage{xspace}
\usepackage{xcolor}
\usepackage[normalem]{ulem}
\usepackage[abs]{overpic}
\usepackage{comment}
\usepackage{dsfont}
\usepackage{slashed}
\usepackage{braket}
\usepackage{graphicx}
\usepackage{booktabs}

\usepackage[dvipsnames]{xcolor}

\allowdisplaybreaks[2]

\def \md {\mathrm{d}}
\newcommand{\sla}[1]{{#1}\!\!\!\slash}
\DeclareRobustCommand\dsout{\bgroup\markoverwith{\color{blue}{\rule[0.4ex]{2pt}{0.8pt}}}\ULon}

\newcommand{\beq}{\begin{equation} }
\newcommand{\eeq}{\end{equation} }

\begin{document}

\preprint{} 
\title{\boldmath
Decoherence in high energy collisions as renormalization group flow
\unboldmath}

\author{Jiayin Gu}
\email{jiayin\_gu@fudan.edu.cn}
\affiliation{Department of Physics and Center for Field Theory and Particle Physics, Fudan University, Shanghai, 200438, China}
\affiliation{Key Laboratory of Nuclear Physics and Ion-beam Application (MOE), Fudan University, Shanghai, 200433, China}

\author{Shi-Jia Lin}
\email{sjlin21@m.fudan.edu.cn}
\affiliation{Department of Physics and Center for Field Theory and Particle Physics, Fudan University, Shanghai, 200438, China}
\affiliation{Department of Physics, University of Chicago, Chicago, IL 60637, USA}

\author{Ding Yu Shao}
\email{dingyu.shao@cern.ch}
\affiliation{Department of Physics and Center for Field Theory and Particle Physics, Fudan University, Shanghai, 200438, China}
\affiliation{Key Laboratory of Nuclear Physics and Ion-beam Application (MOE), Fudan University, Shanghai, 200433, China}
\affiliation{Shanghai Research Center for Theoretical Nuclear Physics, NSFC and Fudan University, Shanghai 200438, China}

\author{Lian-Tao Wang}
\email{liantaow@uchicago.edu}
\affiliation{Department of Physics, University of Chicago, Chicago, IL 60637, USA}
\affiliation{Enrico Fermi Institute, University of Chicago, Chicago, IL 60637, USA}
\affiliation{Kavli Institute for Cosmological Physics, University of Chicago, Chicago, IL 60637, USA}
\affiliation{Leinweber Institute for Theoretical Physics, The University of Chicago, Chicago,
IL 60637, USA}

\author{Si-Xiang Yang}
\email{sxyang@stanford.edu}
\affiliation{Department of Physics and Center for Field Theory and Particle Physics, Fudan University, Shanghai, 200438, China}
\affiliation{Physics Department, Stanford University, Stanford, CA 94305, USA}


\begin{abstract}
The unification of quantum information science and collider physics is opening a new frontier in high-energy experiments, making a systematic understanding of decoherence a critical challenge. We present a framework to systematically compute spin decoherence from final-state radiation by combining soft-collinear effective theory and open quantum system techniques. We demonstrate that the renormalization group (RG) evolution of the final-state spin density matrix constitutes a quantum channel, where the RG flow parameter, rather than time, drives a Markovian loss of quantum information. Our approach incorporates explicit detector resolution parameters, allowing a direct connection between experimental capabilities and the preservation of quantum coherence. Applying this formalism to a fermion pair ($f\bar{f}$) in the high-energy limit with QED-like final-state radiation, we provide the first systematically RG-improved prediction for decoherence as a function of experimental resolution, revealing the underlying decoherence mechanism to be a phase-flip channel. This work establishes an essential theoretical tool for future precision measurements of quantum phenomena in high-energy collisions and offers a new perspective on the interplay between RG flow and decoherence of open quantum systems.

\end{abstract}

\maketitle
{\it Introduction. --}
The study of quantum information in high-energy collider physics is rapidly transitioning from a theoretical curiosity to an experimental reality. Recent breakthroughs, such as the observation of spin entanglement in top-quark pairs~\cite{ATLAS:2023fsd, CMS:2024pts}, have established particle colliders as novel laboratories for studying quantum mechanics at unprecedented energy scales. This progress has spurred a significant theoretical effort to develop new quantum observables and measurement strategies~\cite{Afik:2020onf, Severi:2021cnj, Afik:2022dgh, Aguilar-Saavedra:2022mpg, Aguilar-Saavedra:2022uye, Aguilar-Saavedra:2022wam, Altakach:2022ywa, Ashby-Pickering:2022umy, Barr:2022wyq, Severi:2022qjy, Afik:2022kwm, Aguilar-Saavedra:2023hss, Aoude:2023hxv, Bernal:2023ruk, Bi:2023uop, Cheng:2023qmz, Dong:2023xiw, Ehataht:2023zzt, Fabbri:2023ncz, Fabbrichesi:2023cev, Fabbrichesi:2023jep, Han:2023fci, Ma:2023yvd, Morales:2023gow, Aguilar-Saavedra:2024fig, Barr:2024djo, Cheng:2024btk, Cheng:2024rxi, Fabbrichesi:2024rec, Fabbrichesi:2024wcd, Fabbrichesi:2024xtq, Guo:2024jch, Han:2024ugl, Morales:2024jhj, White:2024nuc, Wu:2024asu, Khan:2020seu,Qian:2020ini,Wu:2024mtj, Wu:2024ovc, Altomonte:2024upf, Fang:2024ple, Du:2024sly, Ruzi:2024cbt, Gabrielli:2024kbz, Maltoni:2024tul, Maltoni:2024csn, Afik:2025grr, Cheng:2025cuv, Cheng:2025zcf, Fabbrichesi:2025ywl, Feleppa:2025clx, Fucilla:2025kit, Goncalves:2025qem, Goncalves:2025xer, Hentschinski:2025pyq, Kutak:2025tsx, Lin:2025eci, Pei:2025non, Qi:2025onf, Ruzi:2025jql, vonKuk:2025kbv, Wu:2025dds,Liu:2025bgw, Han:2025ewp, Aoude:2025jzc, Zhang:2025mmm, Aguilar-Saavedra:2025byk, Ai:2025wnt, Afik:2025ejh}.

A critical challenge for this emerging program is that any realistic quantum system is open. Entangling with the environment inevitably induces decoherence~\cite{Zeh:1970zz, Zurek:1981xq, Zurek:1982ii}. 
In particle physics, decoherence effects have been explored in diverse settings, ranging from flavor oscillations to effective field theories, among which the decoherence mechanism that is particularly relevant in high-energy scattering processes is the emission of unresolved soft and collinear radiation \cite{Burgess:2024heo, Alok:2024amd, Caban:2007je, Carney:2017jut, Carney:2017oxp, Carney:2018ygh, Aoude:2025ovu}. 
While it has been argued that soft emissions preserve spin coherence at leading-order due to soft theorems~\cite{Carney:2017jut}, a systematic framework to quantify decoherence from collinear radiation is still under active development.  Recently, Ref.~\cite{Aoude:2025ovu} studied decoherence from various types of collinear radiation in maximally entangled fermion pairs, identifying Kraus operators with the Altarelli-Parisi splitting functions.  Their fixed-order perturbative analysis already reveals a non-negligible decoherence effect, in particular when the coupling is large or the particles are boosted.  In precisely these regimes, 
the resummation of large logarithms becomes indispensable, and the detector resolution (which determines how well collinear particles can be resolved) also plays a crucial role.  A more comprehensive framework that implements these important effects is thus urgently needed as the field moves towards a more precise entanglement prediction.\footnote{We assume quantum field theory is the underlying description and do not address debates concerning tests of local hidden variable theories at colliders~\cite{Abel:1992kz, Li:2024luk, Bechtle:2025ugc, Abel:2025skj, Low:2025aqq}.}

\begin{figure}[t]
    \centering
    \includegraphics[width=0.9\linewidth]{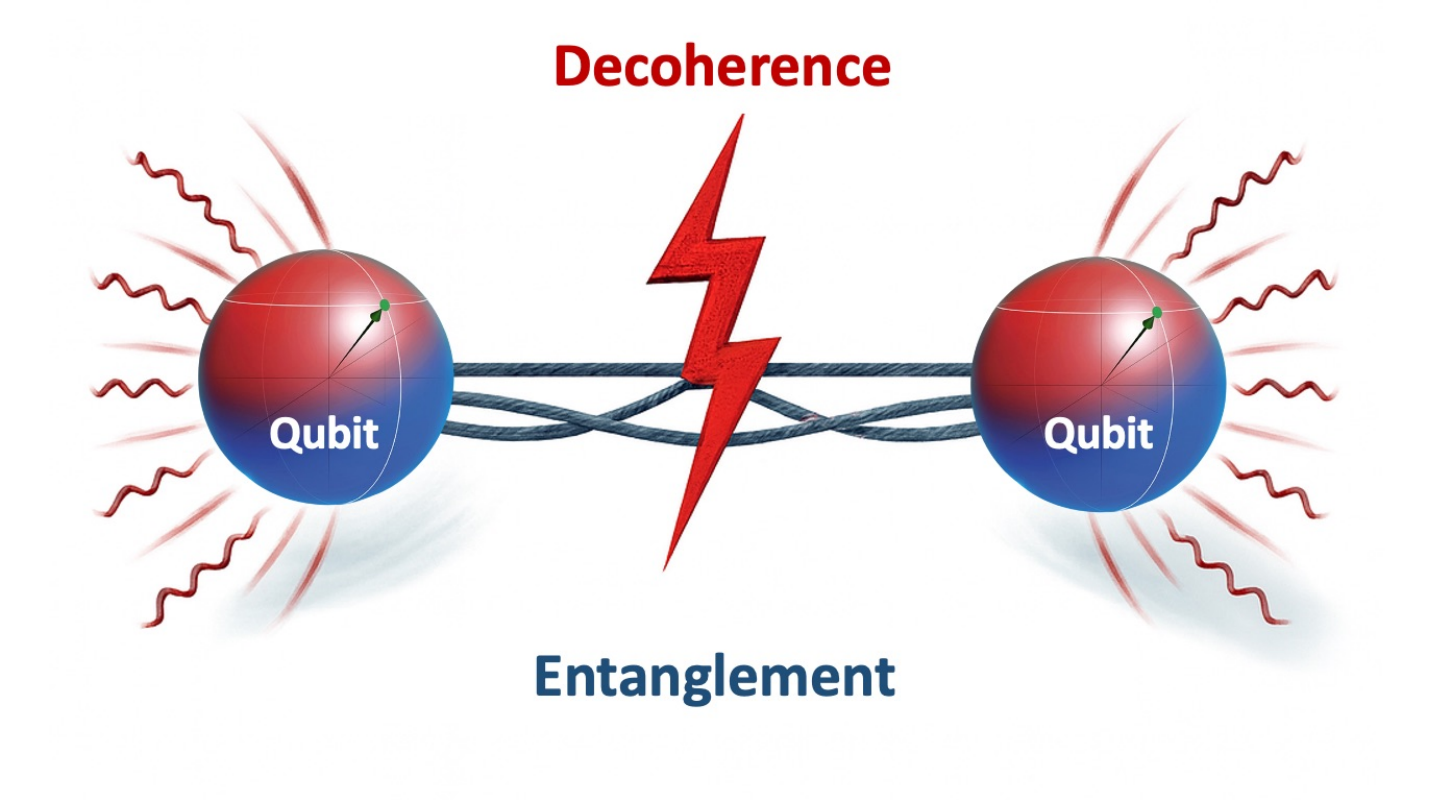}
    \caption{Schematic illustration of spin entanglement generation and loss in two-qubit quantum systems at colliders. The fermion pair is entangled at production but undergoes decoherence due to unresolved soft and collinear emissions.}
    \label{fig:decoherence}
\end{figure}

In this Letter, we provide the first systematic treatment of the spin decoherence in an entangled fermion pair from final-state radiation (FSR) (illustrated in Fig.~\ref{fig:decoherence}) within an effective field theory (EFT). By combining Soft-Collinear Effective Theory (SCET)~\cite{Bauer:2000yr, Bauer:2001ct, Bauer:2001yt, Bauer:2002nz, Beneke:2002ph} with the tools of open quantum systems, we factorize the process into three stages. At short distances, the fermion pair is created, described by a production density matrix. The system then evolves from the production scale to a lower energy scale relevant for the measurement, with the evolution governed by an evolution operator that accounts for decoherence from environmental interactions. Finally, the projection of the evolved state onto the Hilbert space of the observed particles is included in the measurement operator, which provides a direct link to experimental spin observables.

A central result of our work is the demonstration that the renormalization group (RG) evolution of the density matrix constitutes a quantum channel, which governs a Markovian flow of quantum information from the hard interaction scale down to the measurement scale. This provides a novel physical interpretation of RG flow as the engine of decoherence. In addition, our approach makes the notion of ``unresolved radiation'' precise by incorporating explicit detector resolution parameters, allowing for a direct connection between experimental capabilities and the preservation of quantum coherence. 

As a concrete application, we compute the Kraus operators and quantum master equation for a pair of fermions produced in the high energy scattering process, which then undergoes QED-like radiation. Our all-order calculation reveals a measurable suppression of entanglement via an RG equation driven by collinear radiation. This decoherence is driven not by time evolution as in many atomic, molecular, and optical (AMO) systems, but by the renormalization group flow from the hard production energy down to the measurement scale, which often spans many orders of magnitude ({\it e.g.} from TeV to sub-MeV).  Although demonstrated here in QED, the framework is readily generalizable to QCD, which
offers a systematically improvable prediction for entanglement loss in hadronic final states.

{\it Open system and decoherence. --}
To quantify the loss of entanglement, we model the produced fermion pair as an open quantum system. The initial spin state, generated by the short-distance hard scattering, is described by a bipartite density matrix $\hat{\rho}_{\text{hard}}(Q,\mu)$. This matrix includes contributions from the leading-order (LO) process and its virtual corrections with the same spin states. After applying a standard multiplicative renormalization scheme to regularize both ultraviolet (UV) and infrared (IR) divergences, 
$\hat{\rho}_{\text{hard}}$ necessarily depends on the hard scale $Q$ and the renormalization scale $\mu$. In general, it can be decomposed as 
\begin{align}\label{eq:bipartite_density_matrix}
\hat{\rho}_{\text{hard}}(Q,\mu)= \frac{1}{4} \!\left( \hat{I} \!\otimes\! \hat{I} \!+\!  P^+_i\, \hat{\sigma}_i \otimes \hat{I} \!+\!  P^-_j\, \hat{I}\!\otimes\! \hat{\sigma}_j \!+\!  C_{ij}\, \hat{\sigma}_i \!\otimes\! \hat{\sigma}_j \right),
\end{align}
where $\hat{I}$ is the $2\times 2$ identity matrix, $\hat{\sigma}_i$ are the Pauli matrices, and summation over repeated indices is implied. The coefficients $P^{\pm}_i$ and $C_{ij}$ are functions of the kinematic variables. $P^+_i$ and $P^-_j$ denote the components of the polarization vectors of the particle and antiparticle, respectively. The correlation matrix $C_{ij} = \text{Tr}[ \hat{\rho}\, \hat{\sigma}_i \otimes \hat{\sigma}_j ]$ encodes the spin correlations between the two subsystems along directions $i$ and $j$.  

Through soft and collinear emissions, the fermion pair becomes entangled with an environment consisting of unobserved FSR. Tracing over these environmental degrees of freedom induces a non-unitary evolution of the fermion spin state~\cite{Aoude:2025ovu}. This physical process is described by a quantum channel, $\mathcal{E}$, which maps the initial density matrix to a final, mixed state: $\hat{\rho}_{\text{final}} = \mathcal{E}(\hat{\rho}_{\text{hard}})$. According to the Kraus representation theorem~\cite{kraus1983states, Nielsen:2012yss}, any such trace-preserving, completely positive map can be expressed as
\begin{equation}\label{eq:KrausRep}
    \mathcal{E}(\hat{\rho}) = \sum_{\alpha} K_{\alpha} \hat{\rho} K_{\alpha}^\dagger,
\end{equation}
where the Kraus operators $\{K_{\alpha}\}$ satisfy the closure relation $\sum_{\alpha} K_{\alpha}^\dagger K_{\alpha} = \hat{I}$. This formalism provides a rigorous description of the information loss arising from the system-environment interaction, allowing us to precisely calculate the effects of decoherence.

\begin{figure}[t]
    \centering
    \includegraphics[width=\linewidth]{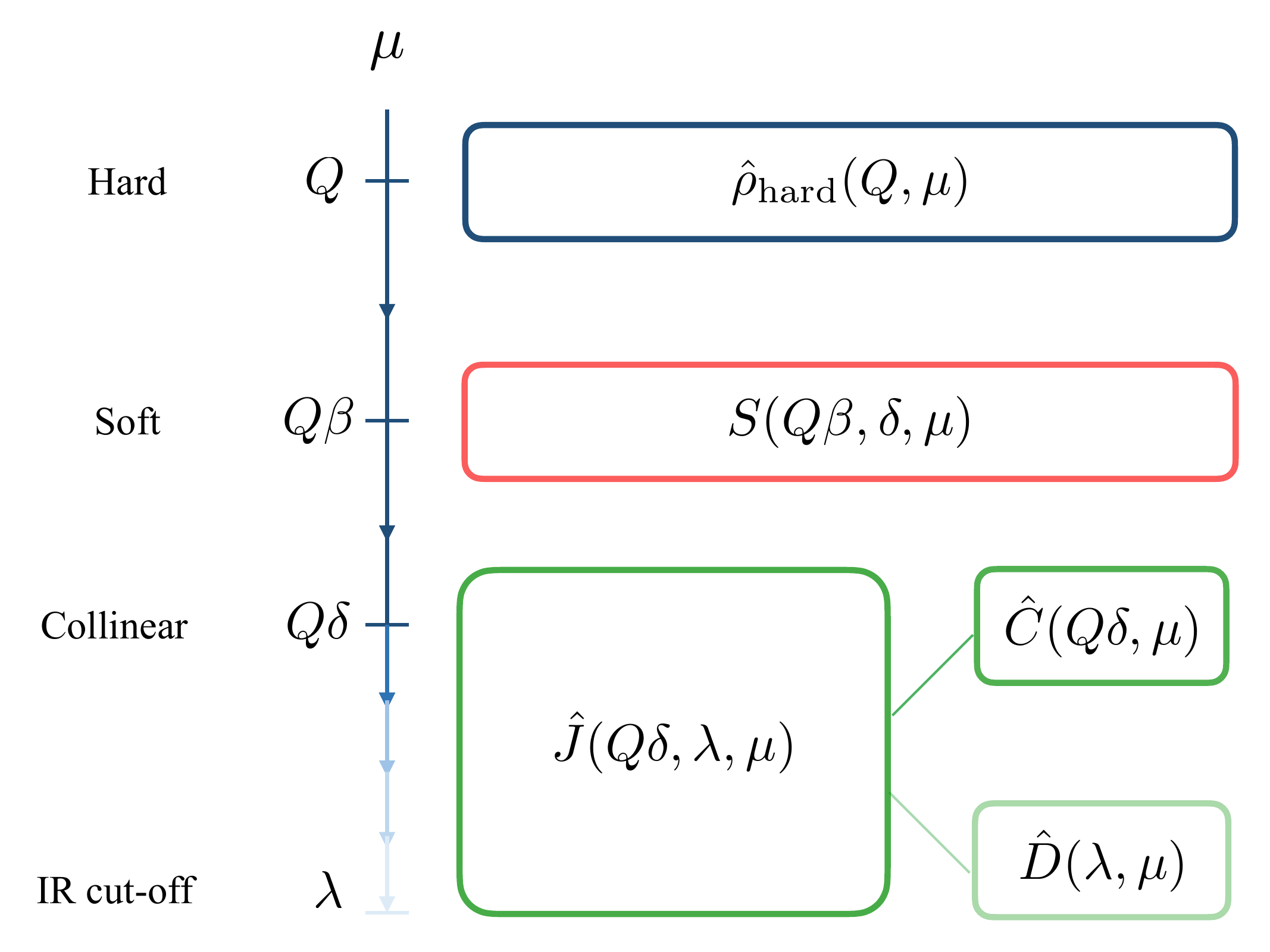}
    \caption{Schematic representation of factorization, the scale separation, and the RG flow in our calculation. Note that we choose $Q\beta>Q\delta$ for illustrative purposes only.} 
    \label{fig:definition} 
\end{figure}

{\it EFT framework. --}
To compute the Kraus operators for the decoherence channel, we use SCET to systematically factorize the dynamics of the fermion pair system from its radiative environment. We define a \textit{fermion jet} in analogy with Sterman-Weinberg jets~\cite{Sterman:1977wj}: an event is classified as a two-fermion final state if the total energy of all radiation outside two cones of half-angle $\delta$ around the fermion momentum axes is less than $Q\beta$, where $\delta$ specifies the angular resolution, and $\beta$ controls the allowed out-of-cone energy. This definition provides a precise physical regulator, isolating the unresolved collinear radiation within the cones as the environment responsible for decoherence.\footnote{Other regulators, such as jet vetoes or event shapes, share similar factorization structures despite their different precise definitions. Hence, we expect they lead to similar results.
}  
This jet definition allows us to apply the factorization theorems of SCET. In the limit $\beta, \delta \ll 1$, the production matrix $\hat{R}$ (from which the spin density matrix is obtained via normalization, $\hat{\rho}=\hat{R}/\operatorname{Tr}[\hat{R}]$) factorizes as\footnote{In this work, we have neglected the factorization structure for non-global logarithms \cite{Dasgupta:2001sh}, since it is irrelevant to the decoherence we are interested in. A complete factorization theorem can be derived via including the multi-Wilson structures \cite{Becher:2015hka}.}
\begin{align}\label{eq:production_density_matrix}
\!\!\hat{R} = \! S(Q\beta, \delta,\mu)\hat J_{\!f}(Q\delta,\lambda,\mu) \hat{R}_{\text{hard}}(Q,\mu)\hat J_{\!\bar{f}}(Q\delta,\lambda,\mu).
\end{align}
Here, $\lambda$ is an infrared regulator for collinear singularities. The factorization and relevant scales are shown in Fig.~\ref{fig:definition}. 
The soft function $S$ accounts for large-angle soft radiation. At the leading power in $\beta$, soft emissions are spin-independent and thus do not induce decoherence~\cite{Low:1954kd, Gell-Mann:1954wra, Lysov:2014csa}. The fragmenting jet operators, $\hat{J}_{f(\bar f)}$, project the hard scattering state onto the Hilbert space of the observed particles, which is defined via the light-cone correlator of fermion fields $\Xi\propto \sum_X\langle 0 | \psi |fX\rangle\langle fX | \bar\psi | 0 \rangle$. This effectively traces over unobserved collinear radiation, and induces decoherence for the entangled spin system. Explicitly, $\hat{J}$ can be decomposed in a basis of spin operators \cite{Jaffe:1996wp, Chen:1994ar}
\begin{align}\label{eq:FJF_operator}
  \!\hat{J}\!=\!\frac{1}{2}\!\left[\mathcal{J}^U\!\hat I\!\otimes\!\hat I\!+\!\mathcal{J}^L\hat{\sigma}_z\!\otimes\!\hat{\sigma}_z \!+\!\mathcal{J}^T(\hat{\sigma}_x\!\otimes\!\hat{\sigma}_x\!+\!\hat{\sigma}_y\!\otimes\!\hat{\sigma}_y)\right],
\end{align}
where the coefficients, $\mathcal{J}^{\mathcal{P}},\mathcal{P}={U,L,T}$, are the unpolarized, longitudinally, and transversely polarized fragmenting jet functions, respectively.  The detailed derivations are provided in the Supplemental Material.

In the limit $Q\delta\gg \lambda$, we perform a further factorization via an operator product expansion to isolate the physics at different scales. The fragmenting jet operators can be decomposed into the product of the matching operator and the fragmentation operator as $\hat{J}(Q\delta,\lambda,\mu)= \hat{C} (Q\delta,\mu)\hat{D}(\lambda,\mu)$.  This allows us to define a scale-dependent effective production matrix, 
\begin{align} \label{eq:reff}
     \!\!\hat{R}_{\text{eff}}(\mu)\equiv\!S(Q\beta, \delta,\mu)\hat{C}_{\!f} (Q\delta,\mu)\hat{R}_{\text{hard}}(Q,\mu)\hat{C}_{\!\bar{f}} (Q\delta,\mu),
\end{align}
which absorbs all components from the hard scale to the factorization scale. Substituting it into Eq.~\ref{eq:production_density_matrix}, the full production matrix is $\hat{R} =\hat{D}_f(\lambda,\mu)\,\hat{R}_{\text{eff}}(\mu)\,\hat{D}_{\bar{f}}(\lambda,\mu)$. 

A key insight of the EFT framework is that its RG consistency governs the evolution of the spin system. Specifically, the requirement that the full production matrix be independent of the arbitrary factorization scale ($\md \hat{R}/ \md \mu=0$) imposes a powerful constraint on the evolution equation of $\hat{R}_{\text{eff}}(\mu)$. Introducing the RG flow parameter, $t\equiv\log(Q\delta/\mu)$, the RG solution of the effective production matrix reads
\begin{align}\label{eq:evolution}
\hat{R}_{\text{eff}}(t)=\hat{U}_f(t,0)\,\hat{R}_{\text{eff}}(0)\,\hat{U}_{\bar{f}}(t,0) \,,
\end{align}
where the evolution operators $\hat{U}_f(t,0)$ and $\hat{U}_{\bar{f}}(t,0)$ are decomposed as the same as Eq.~\eqref{eq:FJF_operator} (see Supplemental Material).  The decomposition coefficients are the polarized evolution functions, 
\begin{align}\label{eq:evol}
  U^{\mathcal{P}}(t,0) = \exp\left(\int_{0}^{t} \md t\, \gamma^{\mathcal{P}}\right),
\end{align}
where $\gamma^{\mathcal{P}} \equiv \frac{\alpha}{\pi} P_{f f}^{\mathcal{P}}$ is the anomalous dimension defined via the first Mellin moment of the Altarelli-Parisi splitting function. Here, we consider only a single type of branching with small-angle emission and ignore the sub-leading effects from the off-diagonal contribution in the flavor space. Eq.~\eqref{eq:evolution} shows the evolution of the effective production matrix as a function of $t$ (starting from $t=0$), during which decoherence occurs. From the evolution operators $\hat{U}_f(t,0)$ and $\hat{U}_{\bar{f}}(t,0)$, we can construct the explicit Kraus operators, $\{K_{\alpha}\}$, for the corresponding quantum channel. This formalism transforms the calculation of decoherence into the one of RG evolution, where the anomalous dimensions directly determine the strength and nature of the information loss. Furthermore, the dynamical map is multiplicative, allowing us to write a differential equation characterizing the open system dynamics, namely the quantum master equation \cite{Nielsen:2012yss, Rivas:2012ugu}.

The final stage of the process is the projection of the evolved spin state onto a definite experimental outcome. The spin-dependent differential cross section is given by the trace of the full production matrix against the measurement projectors for the final-state fermions, $\hat{P}_{f(\bar{f})} = (\hat{I} + \boldsymbol{S}_{f(\bar{f})} \cdot \hat{\boldsymbol{\sigma}})/2$ \cite{Chen:1994ar, Lin:2025eci}, 
\begin{align}\label{eq:sec_pol}
\md\sigma(\boldsymbol{S}_{f},\boldsymbol{S}_{\bar{f}})
\propto \text{Tr}\left[ \hat P_{f} \otimes \hat P_{\bar{f}} \, \hat{R} \right].
\end{align}
By defining spin-dependent measurement operators $\hat{M}_{f(\bar{f})}(\boldsymbol{S}_{f(\bar{f})}) \equiv \hat{D}_{f(\bar{f})} \hat{P}_{f(\bar{f})}$, the cross section takes the form
\begin{align} \label{eq:trace}
\md\sigma(\boldsymbol{S}_{f},\boldsymbol{S}_{\bar{f}}) \propto
\text{Tr}\left[ \hat{M}_f(\boldsymbol{S}_f, t) \, \hat{R}_{\text{eff}}(t) \, \hat{M}_{\bar{f}}(\boldsymbol{S}_{\bar{f}}, t) \right],
\end{align}
where we used the fact that the production matrix is factorized. In Eq.~\eqref{eq:trace},  physics at different scales is cleanly separated: all the effects of decoherence from collinear radiation are encapsulated in the evolution of the effective production matrix, $\hat{R}_{\text{eff}}$, while the IR physics of the final-state projection is contained entirely within the measurement operators, $\hat{M}$.\footnote{In practice, fermion spins are usually reconstructed from their decay products.  This is formally included in $\hat{M}$.  Such reconstructions rely on theoretical assumptions, which we do not address here.}

\begin{figure}[t]
    \centering
    \includegraphics[width=0.95\linewidth]{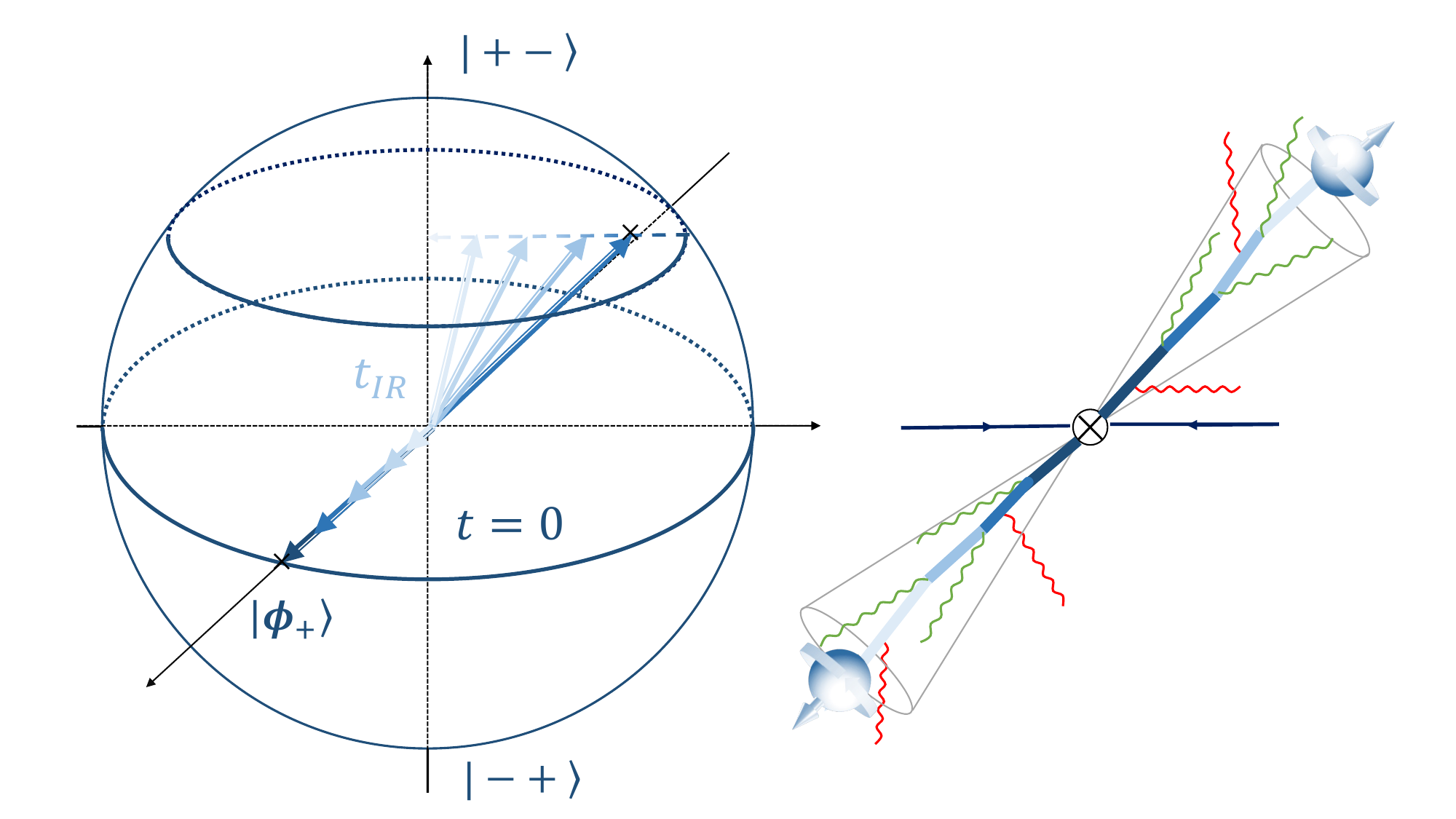}
    \caption{RG evolution as a phase-flip channel. {\bf Left:} Bloch-sphere representation of decoherence 
    in the $\{\ket{+-}, \ket{-+}\}$ subspace, with lighter shading indicating reduced spin correlations.  $\ket{\phi_+} = (\ket{+-}+\ket{-+})/\sqrt{2}$ is a 
    maximally-entangled state (and so are all points on the equator). {\bf Right:} The corresponding picture of spin decoherence driven by collinear photon emissions (green lines).} 
    \label{fig:Dephasing}
\end{figure}

The full process in our EFT framework can be understood through a RG-evolution picture. It unfolds in three stages: At short distances ($t=0$), an entangled fermion pair is created, described by an effective production matrix, $\hat{R}_{\text{eff}}$. The system then evolves to a macroscopic scale $t$ under the influence of an evolution operator, $\hat{U}(t,0)$, which accounts for the decoherence induced by environmental interactions. Finally, the measurement is described by the measurement operator, $\hat{M}$, which projects the evolved state onto the Hilbert space of the observed final-state particles.\footnote{While our analysis focuses on decoherence from FSR, the formalism can be extended to include initial-state radiation (ISR). Within our factorized framework, ISR modifies the initial effective density matrix, $\hat{\rho}_{\text{eff}}(t=0)$, which subsequently evolves due to FSR. A detailed study of ISR effects is left for future work.}

{\it Example. --}
To illustrate our framework, we apply it to the spin states of a fermion pair ($f\bar{f}$) produced at a high energy collider.  Here we do not make any assumption on the hard scattering process that produces the fermion pair, but instead work with the most general initial density matrix, $\hat{\rho}_{\text{eff}}(t=0)$.
The spin state of each fermion evolves by emitting unresolved collinear radiation, which we assume to be described by a QED-like theory with a generic coupling $\alpha \equiv g^2/(4\pi)$.
The hard scale $Q$ is set to the center-of-mass energy of the fermion pair, while the fermion mass serves as the infrared regulator, $\lambda=m$ with $m \ll Q$.  The RG evolution of $\hat{\rho}_{\text{eff}}(t)$ from $t=0$ to $t=\log{(Q\delta/m)}$ defines a quantum channel. 

The form of this channel is dictated by the underlying QED-like interactions in the boosted regime, which conserve helicity but not transverse spin \cite{Barone:2003fy,Jaffe:1991ra} . Working in the helicity basis (see Supplemental Material), this leads to a simple phase-flip channel, whose Kraus operators are $\hat{K}_{(i,j)} = \hat{K}^{f}_i \otimes \hat{K}^{\bar{f}}_j$, with 
\begin{align}\label{eq:Kraus_operators}
    \hat{K}^{f(\bar{f})}_{0} = \sqrt{1 - p^2}\, \hat{I}\,,\quad
    \hat{K}^{f(\bar{f})}_{1} = p\, \hat{\sigma}_3\,,
\end{align}
where $p \equiv \sqrt{(1 - e^{-\frac{\alpha}{2\pi} t})/2}$. Note that in deriving Eq.~\eqref{eq:Kraus_operators} we have ignored the running of $\alpha$, in which case the anomalous dimension $\gamma^{\mathcal{P}}$ in Eq.~\eqref{eq:evol} is independent of $t$. We emphasize that the phase-flip decoherence effect discussed here is dominant in the boosted regime. For massive fermions produced near the threshold, the bit-flip and bit-phase-flip channels would be more important ~\cite{Aoude:2025ovu}, while the phase-flip channel is suppressed.

\begin{figure}[t]
    \centering
    \includegraphics[width=0.95\linewidth]{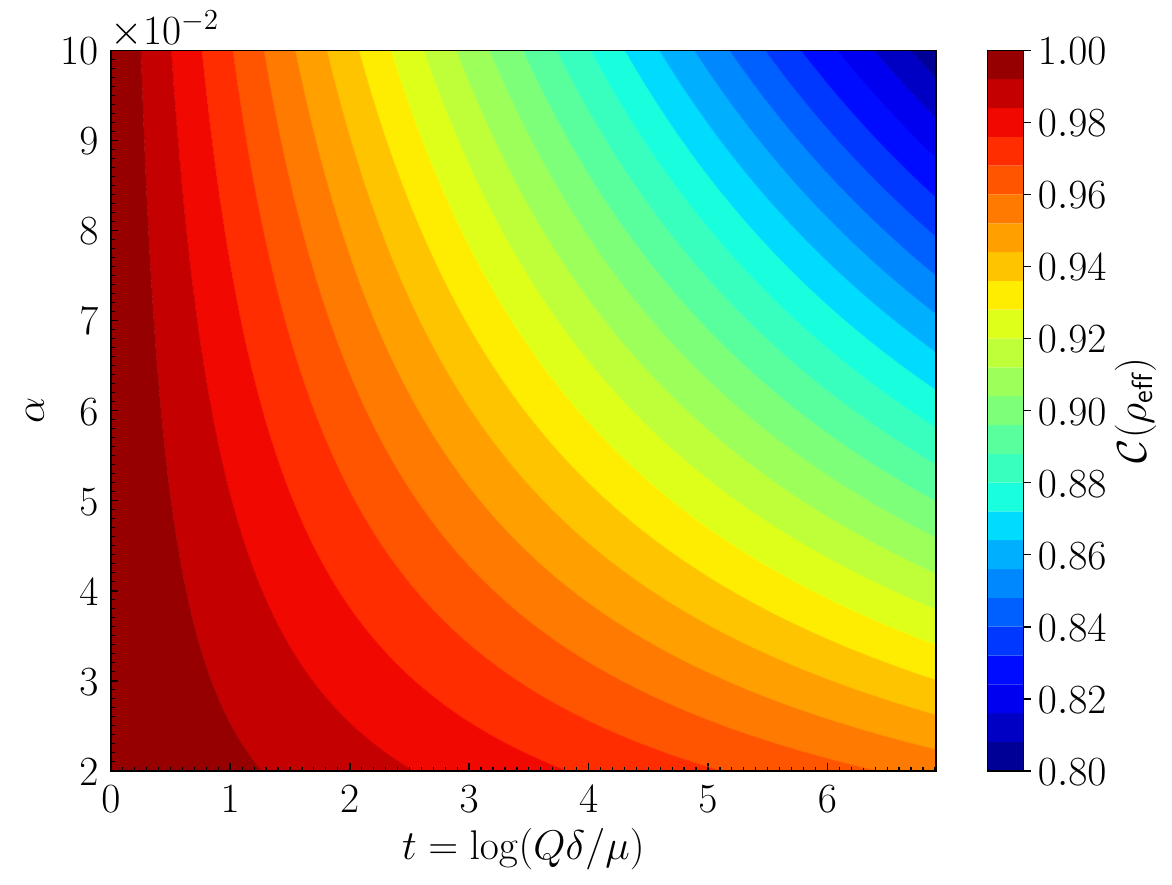}
    \caption{The evolution of concurrence $\mathcal{C}(\rho_{\text{eff}})$ as a function of RG scale $t$ 
    for varying coupling strengths $\alpha$ in the range $[0.02, 0.10]$, assuming $\mathcal{C}(0)=1$ and Eq.~\ref{eq:conevo} holds. Varying $\alpha$ illustrates the general coupling dependence of decoherence, with stronger interactions leading to faster suppression of entanglement.}
    \label{fig:contour}
\end{figure}

The dynamics of this channel are equivalently described by a Lindblad master equation~\cite{Lindblad:1975ef}
\begin{align}
\frac{\md \hat{\rho}_{\text{eff}}}{\md t} 
&= -\frac{\alpha}{2\pi} \hat{\rho}_{\text{eff}} \label{eq:ME} \\
&+ \frac{\alpha}{4\pi} \left[ (\hat{\sigma}_3\otimes\hat{I})\, \hat{\rho}_{\text{eff}}\,(\hat{\sigma}_3\otimes\hat{I})+(\hat{I}\otimes\hat{\sigma}_3)\,\hat{\rho}_{\text{eff}}\,(\hat{I}\otimes\hat{\sigma}_3 )\right], \nonumber
\end{align}  
where the Lindblad jump operators $\hat{L}_1 = \sqrt{\alpha/4\pi}\,\hat{\sigma}_3 \otimes \hat{I}$ and $\hat{L}_2 = \sqrt{\alpha/4\pi}\,\hat{I} \otimes \hat{\sigma}_3$ admit a clear quantum trajectory interpretation: each ``jump" corresponds to an unresolved collinear photon emission from either of the fermion legs, which induces a stochastic phase-flip.  
This leads to an exponential decay of all off-diagonal terms,
\begin{align}
\frac{\hat{\rho}^{ij}_{\text{eff}}(t)}{\hat{\rho}^{ij}_{\text{eff}}(0)}= \begin{cases} 1 & i=j ~~ (\text{diagonal}) \,, \\ e^{-\frac{\alpha}{\pi} t} & i j=14,23,32,41 ~~ (\text{anti-diagonal}) \,, \\ e^{-\frac{\alpha}{2 \pi} t} & \text {else} \,, \end{cases}
\end{align}
where $i,j$ are components in the $\mathcal{H}_f\otimes\mathcal{H}_{\bar{f}}$ spin space. As $t\to \infty$, only the diagonal terms survive and the system becomes a classical mixture of different helicity states. 

For a large class of cases, it is possible to obtain a compact form for the evolution of the concurrence $\mathcal{C}(\hat{\rho}_{\text{eff}}(t))$ (see Supplemental Material for definition), or simply $\mathcal{C}(t)$.  Assuming the $f\bar{f}$ pair are produced with opposite helicities, the spin space effectively reduces to a 2-dimensional subspace formed by $\ket{+-}$ and $\ket{-+}$ (where $\pm$ denotes the sign of the helicity) since the evolution conserves helicity. The corresponding dephasing trajectory on the Bloch sphere in the $\{\ket{+-}, \ket{-+}\}$ subspace always points towards (and is perpendicular to) the $z$-axis, as illustrated in Fig.~\ref{fig:Dephasing}. In this case, $\mathcal{C}(t)$ is simply proportional to the absolute value of the off-diagonal term in the $2\times 2$ density matrix, and is given by
\begin{align} \label{eq:conevo}
\mathcal{C}(t) = \mathcal{C}(0) e^{-\frac{\alpha}{\pi} t} \,,
\end{align}
which is illustrated in Fig.~\ref{fig:contour}.  The same also applies if the $f\bar{f}$ pair are produced with the same helicity.  
These two cases cover many commonly studied processes.  For example, the pair of fermions from Higgs decay $h\to f\bar{f}$ always have the same helicity (and furthermore, they are always in the Bell state $\ket{++}+\ket{--}$\cite{Altakach:2022ywa,Ma:2023yvd})
, while for $e^+e^-\rightarrow Z/\gamma^*\rightarrow f\bar{f}$ at tree-level, the fermion pair always have opposite helicities. 

\begin{figure}[t]
    \centering
    \includegraphics[width=0.95\linewidth]{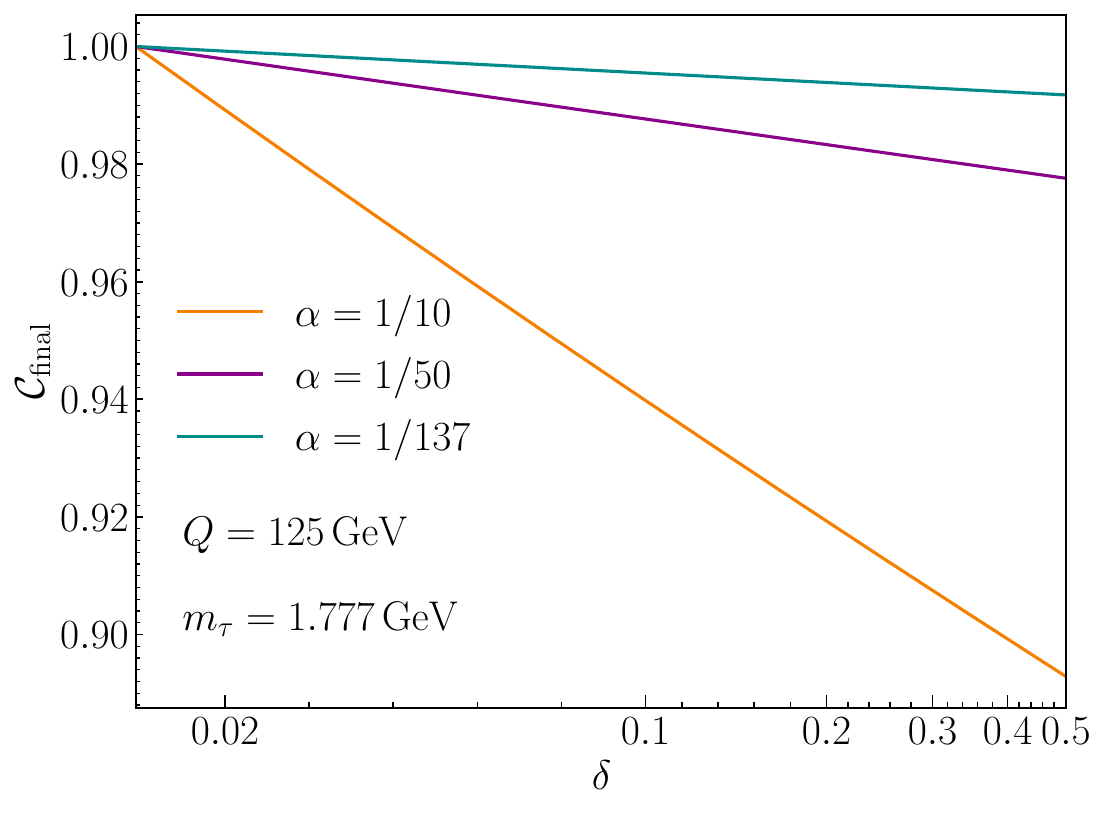}
    \caption{The final concurrence $\mathcal{C}_{\rm final}$ as a function of angular resolution $\delta$, assuming $\mathcal{C}(0)=1$ (initially maximally entangled) and the inequality in Eq.~\ref{eq:con_final} is saturated. 
    As an example, the center-of-mass energy is fixed at $Q = 125~\text{GeV}$, the fermion mass set to $m_\tau \approx 1.777~\text{GeV}$. 
    The three lines correspond to different coupling strengths: $\alpha=1/10$ (orange), $\alpha=1/50$ (dark magenta), and $\alpha=1/137$ (dark cyan). 
    \label{fig:delta}}
\end{figure}

For the most general case, Eq.~\eqref{eq:conevo} instead becomes an inequality
\begin{align} \label{eq:conevo2}
\mathcal{C}(t) \leq \mathcal{C}(0) e^{-\frac{\alpha}{\pi} t} \,,
\end{align}
which can be derived from the factorization law in Ref.~\cite{nat2008} (see also Refs.~\cite{Tiersch:2008zz, Li_2010}) for the concurrence of a two-qubit system that undergoes local noisy channels. Eq.~\eqref{eq:conevo2} leads to the final concurrence at $t=\log(Q\delta/m)$ to satisfy 
\begin{align} \label{eq:con_final}
\mathcal{C}_{\rm final} \leq 
\mathcal{C}(0) \, \left(\frac{Q\delta}{m}\right)^{-\frac{\alpha}{\pi}} \,.  
\end{align}
This expression provides a concrete, testable prediction for entanglement suppression. The crucial feature is the power-law suppression factor, which arises from the resummation of collinear logarithms and directly links the degree of decoherence to the detector's angular resolution ($\delta$) and the fermion mass ($m$). It explicitly shows that as the ability to resolve nearby photons worsens (larger $\delta$), entanglement is more strongly suppressed, a property illustrated in Fig.~\ref{fig:delta}.  

{\it Summary and outlook. --}
In this Letter, we have established the first systematic framework for calculating spin decoherence from FSR in high-energy collisions by unifying SCET with the formalism of open quantum systems. Our central finding is that the RG evolution of the fermion spin state constitutes a quantum channel, where the RG scale evolution drives a Markovian loss of quantum coherence. We provide an explicit calculation 
in QED-like theory, resulting in an analytical formula for entanglement suppression that connects decoherence directly to experimental parameters like angular resolution. This work establishes a systematically improvable EFT framework for calculating decoherence from radiation.

The generality of this framework opens several important avenues for future research. First, the Markovian nature of our result is a direct consequence of the factorization assumption, under which the evolutions of the pair of fermions are independent of each other as well as any additional final state particles in the process.  Non-local interactions ({\it e.g.}, due to QCD confinement) could give rise to important non-Markovian effects which require separate studies. 
Second, our formalism is generalizable to QCD, which will be essential for understanding entanglement in hadronic final states and may offer a novel perspective.  It is also important to include subleading-power soft effects and generalize the formalism to processes such as heavy quark productions for which the effects of fermion masses are important. Ultimately, this line of work will provide the tools to transform the complex environment of a particle collider into a controlled laboratory for studying open quantum systems.

\section{Acknowledgements}

We thank Yoav Afik, Rafael Aoude, Alan J. Barr, Tao Han, Xiaopeng Li, Matthew Low, Fabio Maltoni, Leonardo Satrioni and Pengfei Zhang for helpful discussions and valuable comments on the manuscripts. This work is supported by the National Science Foundations of China under Grant No.~12275052, No.~12147101, No.~12035008 and No.~12375091.  J.G.\ and D.Y.S.\ are also supported by the Innovation Program for Quantum Science and Technology under grant No. 2024ZD0300101. 

\bibliography{Refs}  

\clearpage
\appendix
\onecolumngrid
\setcounter{equation}{0}
\renewcommand{\theequation}{S-\arabic{equation}}

\section*{Supplemental Material}

This Supplemental Material is organized into three parts. 
First, we outline the construction of the density matrix and the definition of concurrence as an entanglement measure. 
Second, we summarize the necessary SCET ingredients, namely the fragmenting jet functions $\mathcal{J}^{\mathcal{P}}$, fragmentation functions $\mathcal{D}^{\mathcal{P}}$, and matching coefficients $\mathscr{C}^{\mathcal{P}}$, which enter the calculation of the density-matrix elements. 
Finally, we present the construction of Kraus operators and the derivation of the master equation, which together describe the evolution of the effective density matrix. 
For concreteness, we illustrate the results with the case where the evolution is governed by QED-like photon radiation.

\section{Production matrix, density matrix, and concurrence}

We always work in the helicity basis throughout this Letter.  The production matrix $\hat{R}$ is constructed from helicity amplitudes as  
\begin{align}
    \hat{R} = \sum_{\lambda_f,\lambda'_f,\lambda_{\bar{f}},\lambda'_{\bar{f}}}
    \mathcal{M}^{\lambda_f\lambda_{\bar{f}}}
    (\mathcal{M}^{\lambda'_f\lambda'_{\bar{f}}})^*
    \ket{\lambda_f\lambda_{\bar{f}}}\bra{\lambda'_f\lambda'_{\bar{f}}}, \label{eq:R_1}
\end{align}
where $\lambda_{f,\bar{f}}$ are defined in each fermion’s helicity frame, with the $z$–axis aligned to its momentum. More explicitly, in the $f\bar{f}$ center-of-mass frame, the coordinate system of $f$ is defined with the $z$-axis along its momentum, $\hat{z}\equiv \boldsymbol{p}_{f}/|\boldsymbol{p}_{f}|$ The $y$-axis is taken perpendicular to the production plane (formed by either the incoming particles, or the heavy particle $f\bar{f}$ decay from), and the $x$-axis completes the right-handed coordinate system, $\hat{x} \equiv \hat{y} \times \hat{z}$. For $\bar{f}$, which moves in the opposite direction, the axes are chosen as $\{\hat{\bar{x}},\hat{\bar{y}},\hat{\bar{z}}\}\equiv\{\hat{x},-\hat{y},-\hat{z}\}$ to maintain a consistent helicity basis for the bipartite system.  Note that our helicity basis is slightly different from the one defined in {\it e.g.} Refs.~\cite{Altakach:2022ywa,Han:2023fci,Cheng:2025cuv}, which uses the same axes for both $f$ and $\bar{f}$.

In the two-qubit space space $\mathcal{H}_f\otimes\mathcal{H}_{\bar{f}}$, $\hat{R}$ admits the Pauli–basis decomposition  
\begin{align} \label{eq:Rdecomp1}
    \hat{R} = \sum_{i,j=0}^3 r_{ij}\, \hat{\sigma}_i \otimes \hat{\sigma}_j ,
\end{align}
where $\hat{\sigma}_{1,2,3}$ are the Pauli matrices and $\hat{\sigma}_0=\hat{I}$, which upon normalization ($\hat{\rho}=\hat{R}/\operatorname{Tr}[\hat{R}]$) gives the density matrix 
\begin{align} \label{eq:bipartite_density_matrix_2}
\hat{\rho} = \frac{1}{4} \!\left( \hat{I} \!\otimes\! \hat{I} \!+\!  P^+_i\, \hat{\sigma}_i \otimes \hat{I} \!+\!  P^-_j\, \hat{I}\!\otimes\! \hat{\sigma}_j \!+\!  C_{ij}\, \hat{\sigma}_i \!\otimes\! \hat{\sigma}_j \right),
\end{align}
as in Eq.~\ref{eq:bipartite_density_matrix}. 
For $\hat{\rho}$ to represent a valid quantum state, it must be a positive semidefinite ({\it i.e.}, all eigenvalues are non-negative), hermitian matrix with unit trace. 

To quantify quantum entanglement in the spin state, we use the concurrence $\mathcal{C}(\rho)$ \cite{Hill:1997pfa}. For a general two-qubit state $\hat{\rho}$, the concurrence is defined as
\begin{align}
\mathcal{C}(\rho) = \max(0, \lambda_1 - \lambda_2 - \lambda_3 - \lambda_4)\,,
\end{align}
where $\lambda_i$ are the square roots of the eigenvalues of the matrix
\begin{align}
\tilde{\rho} = \sqrt{\rho} (\hat{\sigma}_y \otimes \hat{\sigma}_y)\, \rho^*\, (\hat{\sigma}_y \otimes \hat{\sigma}_y)\, \sqrt{\rho},
\end{align}
arranged in decreasing order. A concurrence of $\mathcal{C}=1$ corresponds to a maximally entangled state, while $\mathcal{C}=0$ indicates a fully separable state.

In many cases it is possible to write the concurrence in a simpler form. A useful case for our study is the so-called ``X'' state 
\beq \label{eq:rhox}
\hat{\rho}_X=\left(\begin{array}{cccc}
\rho_{11} & 0 & 0 & \rho_{14} \\
0 & \rho_{22} & \rho_{23} & 0 \\
0 & \rho_{32} & \rho_{33} & 0 \\
\rho_{41} & 0 & 0 & \rho_{44}
\end{array}\right),
\eeq
where the physical conditions on $\hat{\rho}$ imply that $\sum_{i=1}^4\rho_{ii}=1$, $\rho_{ij}^*=\rho_{ji}$, $\rho_{22}\rho_{33} \geq |\rho_{23}|^2$, and $\rho_{11}\rho_{44} \geq |\rho_{14}|^2$. 
Its concurrence can be written as~\cite{Ali:2010unc}
\beq \label{eq:concurX}
\mathcal{C}(\hat{\rho}_X)=2 \max \left( \, 0, \, \left|\rho_{23}\right|-\sqrt{\rho_{11} \rho_{44}}, \, \left|\rho_{14}\right|-\sqrt{\rho_{22} \rho_{33}} \,\right) .
\eeq
In terms of Pauli decomposition in Eq.~\ref{eq:bipartite_density_matrix_2}, $\hat{\rho}$ is in the ``X'' state iff  
\beq
P^\pm_1 = P^\pm_2 =0 \,, \quad\quad 
C_{13}=C_{23}=C_{31}=C_{32}=0 \,.
\eeq

Furthermore, if $f$ and $\bar{f}$ always have opposite helicities (OH), the density matrix (denoted as $\hat{\rho}_{\rm OH}$) reduces to a 2-dimensional subspace formed by $\ket{+-}$ and $\ket{-+}$ (where $\pm$ denotes the sign of the helicity), and its concurrence takes a even simpler form that can be easily deduced from Eq.~\ref{eq:concurX},
\beq \label{eq:rho1}
\hat{\rho}_{\rm OH}=\left(\begin{array}{cccc}
0 & 0 & 0 & 0 \\
0 & \rho_{22} & \rho_{23} & 0 \\
0 & \rho_{32} & \rho_{33} & 0 \\
0 & 0 & 0 & 0
\end{array}\right), \hspace{1cm}
\mathcal{C}(\hat{\rho}_{\rm OH}) = 2\left|\rho_{23}\right| = 2\left|\rho_{32}\right| \,.
\eeq
Similarly, if $f$ and $\bar{f}$ always have same helicities (SH), the density matrix and concurrence are given by
\beq \label{eq:rho2}
\hat{\rho}_{\rm SH} =\left(\begin{array}{cccc}
\rho_{11} & 0 & 0 & \rho_{14} \\
0 & 0 & 0 & 0 \\
0 & 0 & 0 & 0 \\
\rho_{41} & 0 & 0 & \rho_{44}
\end{array}\right),  \hspace{1cm}
\mathcal{C}(\hat{\rho}_{\rm SH}) = 2\left|\rho_{14}\right| = 2\left|\rho_{41}\right| \,.
\eeq
%


\section{SCET ingredients}

We now present the explicit SCET ingredients entering the factorization formula in Eq.~\eqref{eq:production_density_matrix}, using QED as an example. Throughout this section the fermion mass $\lambda=m$ serves as the infrared regulator. We begin with the one-loop expressions for the functions that appear in the factorized production matrix, and then discuss how the renormalization group (RG) evolution determines the effective production matrix $\hat{R}_{\text{eff}}$.

In Eq.~\eqref{eq:production_density_matrix}, the short-distance contribution $\hat{R}_{\text{hard}}$ factorizes into the hard function $H(Q,\mu)$, which encodes virtual corrections, multiplied by the leading-order production matrix $\hat{R}_{\text{LO}}$. In QED, the one-loop results read~\cite{Sterman:1977wj, Becher:2015hka}
\begin{align}\label{eq:SCET_func}
    H(Q,\mu) & =1+\frac{\alpha}{4 \pi}\left[-8\log^2\left(\frac{Q}{\mu}\right)+12\log\left(\frac{Q}{\mu}\right)-16+\frac{7\pi^2}{3}\right], \nonumber\\
    S(Q\beta,\delta,\mu) & =1+\frac{\alpha}{4 \pi}\left[-16\log \delta  \log\left(\frac{Q\beta}{\mu}\right) -8\log^2\delta-\frac{2\pi^2}{3}\right]. 
\end{align}

The fragmenting jet functions $\mathcal{J}^{\mathcal{P}}_{f}$ are defined from the lightcone correlation function $\Xi$:
\begin{align} \label{eq:correlation_matrix}
    &\Xi_{\alpha\beta}=\sum_{X} \!\!\!\!\!\!\!\!\int \, \frac{\md z \, \md t}{2\pi z} 
    e^{i t \bar{n}\cdot P / z} \prod_i\theta\!\left(\delta^2 \!-\! \frac{n\cdot \hat p_{X_i}}{\bar n \cdot\hat p_{X_i}}\right)  \bra{0}\chi_{{n},\alpha}(t\bar{n})\ket{f(P S)\,X}\bra{f(P S)\,X}\bar{\chi}_{{n},\beta}(0)\ket{0}\,
\end{align}
where $f(P,S)$ denotes the identified fermion with momentum $P$ and spin $S$, and $X$ labels unresolved collinear radiation with directions $\hat{p}_{X_i}$.

The unpolarized, longitudinally, and transversely polarized fragmenting jet functions in Eq.~\eqref{eq:FJF_operator} are given by
\begin{align}\label{eq:fragmenting jet functions}
    \mathcal{J}^U_{f}(Q\delta, m) = \mathrm{Tr}\left[ \frac{\slashed{\bar{n}}}{2} \Xi \right],\quad
    \mathcal{J}^L_{f}(Q\delta, m) = \mathrm{Tr}\left[\frac{\slashed{\bar{n}}}{2}\gamma_5  \Xi \right],\quad
    {S}^i_{\perp}\mathcal{J}^T_{f}(Q\delta, m) = \mathrm{Tr}\left[\frac{\slashed{\bar{n}}}{2}\gamma^i_{\perp}\gamma_5 \Xi\right],
\end{align}
with the transverse spin vector satisfying $S_\perp^2=-1$ and $S_\perp\cdot P=0$. These quantities are formulated in SCET, where four-momenta are expressed in light-cone coordinates:
\begin{align}
    p^\mu = (p_+, p_-, p_\perp) = p_+\frac{\bar{n}^\mu}{2} + p_- \frac{n^\mu}{2} + p_\perp^\mu,
\end{align}
with $n^\mu = (1, 0, 0, 1)$ and $\bar{n}^\mu = (1, 0, 0, -1)$. The gauge-invariant collinear fermion field is
\begin{align}
    \chi_n = W_n^\dagger \xi_n, \quad{\rm with}\quad \xi_n=\frac{\sla{n} \sla{\bar n}}{4}\psi_c, 
\end{align}
where $\psi_c$ is the collinear fermion field and $W_n$ is the collinear Wilson line:
\begin{align}\label{eq:Collinear-Wilson-line}
    W_n(x)=\exp \left[i e \int_{-\infty}^0 d s \, \bar{n} \cdot A_c(x+s \bar{n})\right].
\end{align}

Combining real and virtual contributions yields the one-loop polarized fragmenting jet functions,
\begin{align}
    \mathcal{J}^{\mathcal{P}}_{f}(Q\delta,m)
    = 1 + \frac{\alpha}{4\pi}
    \left[4\log^2\left(\frac{Q\delta}{\mu}\right)-6\log\left(\frac{Q\delta}{\mu}\right)-\frac{\pi^2}{6}+7+ j^\mathcal{P}\right],
\end{align}
with $j^U=1$, $j^L=-1$, and $j^T=1-2\log(Q\delta/m)$. From these expressions, one can directly verify the RG consistency of the production matrix $\hat R$ in Eq.~\eqref{eq:production_density_matrix}.

The fragmenting jet operators can be decomposed into the product of a matching operator and a fragmentation operator, $\hat{J}(Q\delta,\lambda,\mu)= \hat{C} (Q\delta,\mu)\hat{D}(\lambda,\mu)$, where $ \hat{C}(Q\delta,\mu)$ and $\hat{D}(\lambda,\mu)$ are defined in the same manner as in Eq.~\eqref{eq:FJF_operator}, with matching coefficients $\mathscr{C}^{\mathcal{P}}$ and polarized fragmentation functions $ \mathcal{D}^{\mathcal{P}}$ replacing the polarized fragmenting jet functions. The fragmentation functions are defined analogously to the fragmenting jet functions but without the cone constraint. At one loop, the first Mellin moments of the fragmentation functions are
\begin{align}
    \mathcal{D}^{\mathcal{P}}_{f}= 1 + \frac{\alpha}{4\pi}d^\mathcal{P},\qquad {\rm with}\quad
    d^U=0,\quad d^L=-2,\quad d^T=-2\log(\mu/m).
\end{align}
The one-loop matching coefficients then follow from $\mathscr{C}^{\mathcal{P}(1)}_{f}= \mathcal{J}^{\mathcal{P}(1)}_{f}- \mathcal{D}^{\mathcal{P}(1)}_{f}$, yielding
\begin{align}
     & \mathscr{C}^{U}_{f}= \mathscr{C}^{L}_{f}
    = 1 + \frac{\alpha}{4\pi}
    \left[4\log^2\left(\frac{Q\delta}{\mu}\right)-6\log\left(\frac{Q\delta}{\mu}\right)-\frac{\pi^2}{6}+8\right],\\
    & \mathscr{C}^{T}_{f}
    = 1 + \frac{\alpha}{4\pi}
    \left[4\log^2\left(\frac{Q\delta}{\mu}\right)-8\log\left(\frac{Q\delta}{\mu}\right)-\frac{\pi^2}{6}+8\right].
\end{align}

We now turn to the RG evolution. For $\mu>Q\delta$, the evolution is spin–independent and thus does not affect decoherence in spin space. By contrast, for $\mu\lesssim Q\delta$ the DGLAP evolution arises from collinear emission and becomes spin–dependent. The fragmentation functions evolve according to the DGLAP equation in Mellin space:
\begin{align}
    \frac{\md}{\md\ln\mu} \mathcal{D}^\mathcal{P}_{f}(m,\mu)
    = \gamma^{\mathcal{P}}\, \mathcal{D}^\mathcal{P}_{f}(m,\mu),
\end{align}
with anomalous dimensions
\begin{align}
    \gamma^U=\gamma^L=0,\qquad \gamma^T=-\frac{\alpha}{2\pi}.
\end{align}
This implies the evolution functions $U^U=U^L=1$ and $U^T=e^{-\Gamma t}$ with $\Gamma=\alpha/(2\pi)$. As a consequence of RG consistency, the effective production matrix $\hat{R}_{\rm eff}(\mu)$ evolves according to Eq.~\eqref{eq:evolution}.


\section{Construction of Kraus operators and master equation}

The evolution of the effective production matrix $\hat{R}_{\rm eff}(t)$ is given by Eq.~\ref{eq:evolution}.  More explicitly, with all the spin indices, we have
\beq\label{eq:evolution2}
\hat{R}_{\text{eff}}(t)_{\alpha\bar{\alpha} ,\beta\bar{\beta}}=\hat{U}_f(t,0)_{\beta_0\alpha,\alpha_0\beta} \,\hat{U}_{\bar{f}}(t,0)_{\bar{\beta}_0\bar{\alpha},\bar{\alpha}_0\bar{\beta}} \,\hat{R}_{\text{eff}}(0)_{\alpha_0\bar{\alpha}_0,\beta_0\bar{\beta}_0}  \,,
\eeq
where $\alpha,\beta$ ($\bar{\alpha},\bar{\beta}$) are the spin indices for $f$ ($\bar{f}$) and a summation over repeated spin indices is implied.  
Note that, as in Eq.~\ref{eq:Rdecomp1}, a general production matrix can be decomposed in the spin space $\mathcal{H}_f\otimes\mathcal{H}_{\bar{f}}$ as 
\beq
\hat{R}_{ \alpha \bar{\alpha}, \beta \bar{\beta}}=\sum_{i, j=0}^3 \gamma_{i j} \, \hat{\sigma}^i_{\alpha \beta} \,\hat{\sigma}^j_{\bar{\alpha} \bar{\beta}} \,.
\eeq
The initial spins are labeled with an additional subscript $0$ (corresponding to $t=0$) and are summed over in Eq.~\ref{eq:evolution2}.  %
We will now drop the label $(t,0)$ in the evolution operators $\hat{U}_f$ and $\hat{U}_{\bar{f}}$ since the information is already contained in the spin indices. 

The evolution operator $\hat{U}_f$ can be parameterized as
\begin{align}
(\hat{U}_f)_{\beta_0\alpha,\alpha_0\beta} & =\frac{1}{2} \sum_{m=0}^3 U_f^m \hat{\sigma}_{\beta_0\alpha_0 }^m \hat{\sigma}_{\alpha\beta }^m  \nonumber \\ 
& =\frac{1}{2} \sum_{m=0}^3 \tilde{U}_f^m \hat{\sigma}_{\alpha \alpha_0}^m \hat{\sigma}_{\beta_0 \beta}^m \,.
\end{align}
The first line is the  
conventional parameterization  \cite{Jaffe:1996wp, Chen:1994ar}, where the real coefficients $U^m_f$ are the  
unpolarized, transversely and longitudinally polarized evolution functions,  
$U_f^0=U^U_f,~U_f^{1,2}=U^T_f$ and $U_f^3=U^L_f$.  In the second line, we have rearranged the spin indices in order to write the results in the Kraus representation as in Eq.~\ref{eq:Kraus_operators}.  Note the ordering of spin indices is reversed for $\hat{\sigma}^m_{\beta_0\beta}$ since it acts on $\langle \beta_0 |$.  It is straightforward to work out the new coefficients $\tilde{U}_f^m$, which are 
\begin{align}
    \tilde{U}^0_f=&\frac{1}{2}(U^U_f+U^L_f+2U^T_f) \,, \quad\quad
    \tilde{U}^{1}_f=\tilde{U}^{2}_f=\frac{1}{2}(U^U_f-U^L_f)\,, \quad\quad
    \tilde{U}^3_f=\frac{1}{2}(U^U_f+U^L_f-2U^T_f) \,.
\end{align}
The evolution operator $\hat{U}_{\bar{f}}$ can be parameterized analogously.

Now we can write Eq.~\ref{eq:evolution2} as
\beq\label{eq:evolution3}
\hat{R}_{\text{eff}}(t)_{\alpha\bar{\alpha} ,\beta\bar{\beta}} =
\frac{1}{2} \tilde{U}_f^m \hat{\sigma}_{\alpha \alpha_0}^m \hat{\sigma}_{\beta_0 \beta}^m \, 
\frac{1}{2} \tilde{U}_{\bar{f}}^n \hat{\sigma}_{\bar{\alpha}\bar{\alpha}_0}^n \hat{\sigma}_{\bar{\beta}_0 \bar{\beta}}^n \, 
\hat{R}_{\text{eff}}(0)_{\alpha_0\bar{\alpha}_0 ,\beta_0\bar{\beta}_0} \,,
\eeq
where the summation over $m,n=0,1,2,3$ is implied.  Eq.~\ref{eq:evolution3} can be written in a matrix form as
\beq
\hat{R}_{\text{eff}}(t) = \frac{1}{4} \tilde{U}_f^m \tilde{U}_{\bar{f}}^n\left(\hat{\sigma}^m \otimes \hat{\sigma}^n\right) \hat{R}_{\text{eff}}(0) \left(\hat{\sigma}^m \otimes \hat{\sigma}^n\right)^\dagger \,,
\eeq
where we have also used the fact that the Pauli matrices are Hermitian.  

The effective density matrix $\hat{\rho}_{\text{eff}}(t)$ can be obtained simply with a normalization 
\begin{align}
    \hat{\rho}_{\text{eff}}(t)=\frac{1}{\text{Tr}[\hat{R}_{\text{eff}}(t)]}\hat{R}_{\text{eff}}(t) \,,
\end{align}
where
\begin{align}
    \text{Tr}[\hat{R}_{\text{eff}}(t)]=\text{Tr}[\hat{R}_{\text{eff}}(0)]\left(\sum_{m}\!\tilde{U}_f^m\right)\left(\sum_{n}\!\tilde{U}_{\bar{f}}^n\right)=\text{Tr}[\hat{R}_{\text{eff}}(0)]\left(2\,U^U_f\right)\left(2\,U^U_{\bar{f}}\right).
\end{align}
Therefore, we have
\begin{align}
    \hat{\rho}_{\text{eff}}(t) = \sum_{m,n}\hat{K}_{(m,n)}\,\hat{\rho}_{\text{eff}}(0)\,\hat{K}^\dagger_{(m,n)} \,,
\end{align}
%
%
where the Kraus operators are
\begin{align}
\hat{K}_{(m,n)}=\hat{K}^{f}_m\otimes\hat{K}^{\bar f}_n=\frac{\sqrt{\tilde{U}^m_{f}\tilde{U}^n_{\bar{f}}}}{2
\,\sqrt{U^U_{f}U^U_{\bar{f}}}}\,\hat{\sigma}_m\otimes\hat{\sigma}_n\,.
\end{align}
The Kraus operators are also automatically normalized as
\begin{align}\label{eq:krausnorm}
\sum_{m,n}\hat{K}_{(m,n)}\hat{K}^\dagger_{(m,n)}=\frac{\left(\sum_{m}\!\tilde{U}_f^m\right)\left(\sum_{n}\!\tilde{U}_{\bar{f}}^n\right)}{{2U_f^U2U^U_{\bar f}}}{}\hat{I}\otimes\hat{I}=\hat{I}\otimes\hat{I} \,.
\end{align}

We can plug in the QED results obtained in the previous section: $U^U=U^L=1,\,U^T=e^{-\Gamma t}$ to get
\begin{align}
    \tilde{U}^0_{f(\bar f)}=2(1-p^2),\,\tilde{U}^3_{f(\bar f)}=2p^2,\,\tilde{U}^{1}_{f(\bar f)}=\tilde{U}^{2}_{f(\bar f)}=0 \,,
\end{align}
where $p = \sqrt{(1 - e^{-\Gamma t})/2}$ and $\Gamma = \alpha/(2\pi)$. As shown in the main paper, the Kraus operators induced by QED photon radiation are thus
$\hat{K}_{(i,j)} = \hat{K}^{f}_i \otimes \hat{K}^{\bar{f}}_j$, with
\begin{align}
    \hat{K}^{f(\bar{f})}_{0} = \sqrt{1 - p^2}\, \hat{I}\,,\quad
    \hat{K}^{f(\bar{f})}_{1} = p\, \hat{\sigma}_3\,.
\end{align}

Next, we will derive the master equation. In the derivation of Kraus operators, we only considered the mapping from $t=0$ to a later RG scale $\hat{U}(t,0)$. However, the mapping is defined in any given RG scale interval and is multiplicative 
\begin{align}
    \hat{U}(t_1,t_2)=\hat{U}(t_1,t_0)\hat{U}(t_0,t_2) \,.
\end{align}
This is because of the Markovian assumption of the evolution equations, which can be seen from the definition of the evolution functions
\begin{align}
    U^\mathcal{P}(t_1,t_2)=\exp\left(\int_{t_1}^{t_2}{\md t}\,\gamma^{\mathcal{P}}\right).
\end{align}
The Markovian property of the evolution allows us to write down the master equation. It is usually easier to derive the master equation when the Kraus operators are explicitly known, so here we make use of the QED case as an example. The effective density matrices at $t$ and $t+\text{d}t$ are related, up to $\mathcal{O}(\text{d}t)$, as
\begin{align} 
\rho_{\text{eff}}(t+\text{d}t)=  &\sum_{m,n}\hat{K}_{(m,n)}(t+\text{d}t,t)\,\hat{\rho}_{\text{eff}}(t)\,\hat{K}^\dagger_{(m,n)} (t+\text{d}t,t) \nonumber\\
=&\left( 1-\frac{\alpha}{2\pi}\,\text{d}t\right) \rho_{\text{eff}} (t)+ \frac{1}{2}\frac{\alpha}{2\pi}\,\text{d}t \left[ (\hat{\sigma}_3\otimes\hat{I})\, \rho_{\text{eff}}(t)\,(\hat{\sigma}_3\otimes\hat{I})+(\hat{I}\otimes\hat{\sigma}_3)\,\rho_{\text{eff}}(t)\,(\hat{I}\otimes\hat{\sigma}_3 )\right]+\mathcal{O}(\md t^2) \,.
\label{eq:mastdt}
\end{align}
Therefore, the master equation can be written as in Eq.~\ref{eq:ME}. 

For a given initial (effective) density matrix $\hat{\rho}_{\text{eff}}(0)$, the density matrix $\hat{\rho}_{\text{eff}}(t)$ can be obtained by either solving Eq.~\ref{eq:ME} or directly applying the Kraus operators in Eq.~\ref{eq:Kraus_operators}. 
For a most general density matrix, one obtain 
\beq
\sum_{i,j} K_{(i,j)}
\left(\begin{array}{cccc}
\rho_{11} & \rho_{12} & \rho_{13} & \rho_{14} \\
\rho_{21} & \rho_{22} & \rho_{23} & \rho_{24} \\
\rho_{31} & \rho_{32} & \rho_{33} & \rho_{34} \\
\rho_{41} & \rho_{42} & \rho_{43} & \rho_{44}
\end{array}\right)
K_{(i,j)}^\dagger = 
\left(\begin{array}{cccc}
\rho_{11} & 0 & 0 & 0 \\
0 & \rho_{22} & 0 & 0 \\
0 & 0 & \rho_{33} & 0 \\
0 & 0 & 0 & \rho_{44}
\end{array}\right)  
+ e^{-\frac{\alpha}{2\pi}t}
\left(\begin{array}{cccc}
0 & \rho_{12} & \rho_{13} & 0 \\
\rho_{21} & 0& 0 & \rho_{24} \\
\rho_{31} & 0 & 0 & \rho_{34} \\
0 & \rho_{42} & \rho_{43} & 0
\end{array}\right) 
+ e^{-\frac{\alpha}{\pi}t}
\left(\begin{array}{cccc}
0 & 0 & 0 & \rho_{14} \\
0 & 0& \rho_{23} & 0 \\
0 & \rho_{32} & 0 & 0 \\
\rho_{41} & 0 & 0 & 0
\end{array}\right) \,,
\eeq
where all off-diagonal terms exhibit exponential decays. This is the direct consequence of the phase-flip channel that acts locally on each fermion.  The concurrence satisfies the inequality in Eq.~\ref{eq:conevo2},
\begin{align} \label{eq:conevo2_sup}
\mathcal{C}(t) \leq \mathcal{C}(0) e^{-\frac{\alpha}{\pi} t} \,.
\end{align}

If $\hat{\rho}_{\text{eff}}(0)$ is described by the ``X'' state in Eq.~\ref{eq:rhox}, it stays as an ``X'' state under evolution and the concurrence at $t$ is given by
\beq
\mathcal{C}(t)=2 \max \left( \, 0, \, e^{-\frac{\alpha}{\pi}t} \left|\rho_{23}\right|-\sqrt{\rho_{11} \rho_{44}}, \, e^{-\frac{\alpha}{\pi}t} \left|\rho_{14}\right|-\sqrt{\rho_{22} \rho_{33}} \,\right) \,,
\eeq
which also clearly satisfies the inequality Eq.~\ref{eq:conevo2}. 
Furthermore, we can easily see that the inequality is saturated if $\hat{\rho}_{\text{eff}}(0)$ is either in the opposite-helicity form of Eq.~\ref{eq:rho1} or the same-helicity form of Eq.~\ref{eq:rho2}. 

\end{document}